\documentclass{article}
\usepackage{graphicx}
\usepackage{epsfig}
\usepackage{amsfonts}
\usepackage{enumerate} 
 \usepackage{amssymb}
\usepackage{cite}

\date{\today}

\newcommand{\insertplot}[5]{\begin{figure}
 \hfill\hbox to 0.05in{\vbox to #5in{\vfill
 \inputplot{#1}{#4}{#5}}\hfill}
 \hfill\vspace{-.1in}
 \caption{#2}\label{#3}
 \end{figure}}
 \newcommand{\inputplot}[3]{
 \special{ps: plotfile #1}
}
\newcounter{fig}

\newcommand{\ee}{\end{equation}}
\newcommand{\eea}{\end{eqnarray}}
\newcommand{\be}{\begin{equation}}
\newcommand{\bea}{\begin{eqnarray}}

\begin{document}
\begin{center}

{\LARGE \bf 
Spherical electro-vacuum  black holes 
\vspace{0.2cm} \\    with resonant, scalar $Q$-hair}
\vspace{0.8cm}
\\
{{\bf 
Carlos A. R. Herdeiro and 
Eugen Radu
}
\vspace{0.3cm}
\\
$^{\ddagger }${\small Departamento de Matem\'atica da Universidade de Aveiro and } \\ {\small  Centre for Research and Development  in Mathematics and Applications (CIDMA),} \\ {\small    Campus de Santiago, 3810-183 Aveiro, Portugal}
}
\vspace{0.3cm}
\end{center}

\date{September 2019}

\begin{abstract}   
The asymptotically flat, spherical, electro-vacuum black holes (BHs) are shown to support static, spherical configurations  of a gauged, self-interacting, scalar field,  minimally coupled to the geometry. 
Considering a $Q$-ball type potential for the scalar field,
 we dub these configurations \textit{$Q$-clouds}, in the test field approximation. The clouds exist under a \textit{resonance condition}, at the threshold of (charged) superradiance. This is similar to the stationary clouds supported by Kerr BHs, which exist for a \textit{synchronisation condition}, at the threshold of (rotational) superradiance. In contrast with the rotating case, however, $Q$-clouds require the scalar field to be massive \textit{and} self-interacting; no similar clouds exist for massive but free scalar fields. First, considering a decoupling limit, we construct $Q$-clouds around Schwarzschild and Reissner-Nordstr\"om BHs, showing there is always a mass gap. Then, we make the $Q$-clouds backreact, and construct fully non-linear solutions of the Einstein-Maxwell-gauged scalar system describing spherical, charged BHs with resonant, scalar \textit{$Q$-hair}.  Amongst other properties, we observe there is non-uniqueness of charged BHs in this model 
and the $Q$-hairy BHs can be entropically preferred over Reissner-Nordstr\"om, for the same charge to mass ratio; some $Q$-hairy BH solutions can be overcharged.
 We also discuss how some well known no-hair theorems in the literature, applying to electro-vacuum plus minimally coupled scalar fields, are circumvented by this new type of BHs.
\end{abstract}

\tableofcontents

\section{Introduction}
The last few years have brought considerable insight on the interaction between black holes (BHs) and scalar fields, both in General Relativity (GR) and modified gravity. Let us focus on GR, thus considering only scalar fields minimally coupled to the geometry.  A surprising result is that asymptotically flat rotating BHs can support non-trivial configurations of a massive, free scalar field, when a certain \textit{synchronisation} condition holds. Consider the paradigmatic Kerr BH~\cite{Kerr:1963ud}. As a test field, stationary scalar \textit{clouds} exist, obtained by solving the Klein-Gordon equation on the Kerr geometry~\cite{Hod:2012px}. Beyond the linear approximation, these clouds can be made to backreact on the Kerr geometry, becoming BH \textit{hair}. Here, BH ``hair" signifies macroscopic degrees of freedom not associated to a Gauss law. Thus a new sort of BH emerges~\cite{Herdeiro:2014goa}, defying the paradigm that ``BHs have no hair"~\cite{Ruffini:1971bza}, even in GR. In this case, moreover, the hairy BHs bifurcate from the Kerr solution. In other words, the scalar hair can be arbitrarily small.

Synchronised hair can also be added to Kerr-Newman BHs~\cite{Delgado:2016jxq} and spinning BHs in higher dimensional vacuum GR~\cite{Brihaye:2014nba,Herdeiro:2015kha,Herdeiro:2017oyt}. The case of Myers-Perry BHs~\cite{Myers:1986un} shows, however, a key difference with respect to the Kerr case. The hairy generalisations of Myers-Perry BHs do not bifurcate from the vacuum solutions; there is a \textit{mass gap}. A possible interpretation of this difference is given by the phenomenon of \textit{superradiance}~\cite{Brito:2015oca}. A massive scalar field can trigger a superradiant instability of Kerr BHs. In between the stable and unstable modes there are \textit{zero modes}. In fact, the synchronisation condition is the zero mode condition. The hairy generalisations of the vacuum Kerr solution bifurcate from these zero modes. On the other hand, a massive scalar field cannot trigger superradiant instabilities of Myers-Perry BHs. Although superradiant \textit{scattering} can exist, and extract rotational energy from Myers-Perry BHs, no superradiant \textit{bound states} exist, which are the modes triggering the instability. Thus, there are no zero modes. As such, even though  synchronisation provides a mechanism to equilibrate a scalar configuration with a BH horizon -- the scalar flux through the horizon vanishes~\cite{Herdeiro:2015gia} --, the hairy BHs do not bifurcate from the vacuum Myers-Perry solutions due to the absence of zero modes.

Heuristically, the difference between the Kerr and Myers-Perry case can be associated to balance of the (long range behaviour of the) competing forces allowing, or not, the existence of bound states. In four spacetime dimensions, the attractive gravitational potential decays as $1/r$ whereas the repulsive centrifugal potential decays as $1/r^2$. This creates a potential well allowing for bound states, in particular for superradiant states. Thus it allows the existence of superradiant zero modes. On the other hand, in five spacetime dimensions, both the attractive gravitational and the repulsive centrifugal potentials decay as $1/r^2$, preventing the existence of a potential well, and, as such, of bound states for scalar modes. 

The Reissner-Nordstr\"om (RN) BH of electro-vacuum also allows a phenomenon akin to rotational superradiance, often called \textit{charged superradiance}~\cite{Bekenstein:1973mi}. This amounts to the ability of a \textit{charged} bosonic field to extract Coulomb (rather than rotational) energy from the BH. In this case, as in the higher dimensional Myers-Perry case, the competing forces do not favour the creation of a potential well. Indeed, the attractive gravitational potential and the repulsive electrostatic potential both decay as $1/r$. As such, a charged bosonic field can superradiantly scatter from a RN BH, but no superradiant bound states exist and, consequently, no zero modes. Zero modes, in this case, obey what we shall call a \textit{resonance condition}, $cf.$~(\ref{condition}) below. A formal proof of the absence of superradiant bound states on RN was given by Hod~\cite{Hod:2013eea,Hod:2013nn}.

The absence of zero modes for RN does not perclude the existence of hairy extensions of RN BHs under the resonance condition, having a mass gap with RN, similarly to what occurs with the hairy extensions of Myers-Perry BHs. In fact, such hairy BHs exist and the purpose of this paper is to report them.

\medskip
A charged (or \textit{gauged}) scalar field, with a positive potential, minimally coupled to electro-vacuum ($cf.$~(\ref{action}) below), is a model that has been considered in several contexts. The only spherical BH solution known in this model, up to now, is the RN family. In fact, a theorem by Mayo and Bekeinstein~\cite{Mayo:1996mv} is often invoked as proof no other BHs can exist. As we shall see below, however, this is not the case, and the theorem  can be circumvented. The model, moreover, possesses particle-like solutions, known as \textit{charged (or gauged) boson stars}, a generalisation of the usual (uncharged)  boson stars - see~\cite{Schunck:2003kk,Liebling:2012fv} for reviews. Mini-gauged boson stars, for which the scalar potential includes only a mass term, have been discussed in~\cite{Jetzer:1989av,Jetzer:1993nk,Pugliese:2013gsa}. These are gravitating solitons supported by the non-linearities of GR, which trivialise in the flat spacetime limit. A different class of gauged boson stars exists if one considers a sufficiently non-linear scalar field potential. In particular, gauged boson stars with a $Q$-ball potential, $cf.$~(\ref{potential}) below, have been discussed in~\cite{Prikas:2002ij,Brihaye:2014gua}. Due to this potential, the solutions do not trivialise in the flat spacetime limit, wherein they became gauged $Q$-balls, charged generalisations of the flat spacetime scalar solitons named $Q$-balls by Coleman~\cite{Coleman:1985ki}.

The main result in this paper is that electro-vacuum minimally coupled to a gauged scalar field with a positive potential admits new spherical BH solutions with scalar hair, under the aforementioned resonance condition, but self-interactions are \textit{mandatory} in the scalar potential. We shall construct explicit examples with a $Q$-ball potential, naming the new BHs as having \textit{$Q$-hair}. That the $Q$-hairy BHs are an extension of RN BHs can be argue as follows. In a certain decoupling limit, the field equations reduced to a test gauged scalar field on a fixed RN background. Non-trivial solutions can be found, that we name $Q$-clouds, following~\cite{Herdeiro:2014pka}. The existence of these solutions was recently pointed out in~\cite{Hong:2019mcj}. The hairy BHs can be seen as the backreaction of these $Q$-clouds. Interestingly, since the resonance condition can be taken in a gauge wherein the electrostatic potential vanishes at the event horizon, in a different decoupling limit, gauged $Q$-clouds can be found on a fixed Schwarzschild BH background. So our hairy BH solutions can also be seen as a hairy extension of Schwarzschild BHs. 

\medskip
This paper is organised as follows. In Section~\ref{sec2} we present the Einstein-Maxwell-scalar model, the field equations, the ansatz for the fields and discuss a virial identity. In Section~\ref{apform} we discuss the boundary behaviour of the different fields, in particular introducing the resonance condition. In Section~\ref{sec4} we discuss gauge fixing and some quantities of interest for the analysis of the following results. In Section~\ref{sec5} we address how the Mayo-Bekenstein theorem is circumvented, as well as other no-go theorems, namely by Pe\~na and Sudarsky~\cite{Pena:1997cy} and Hod~\cite{Hod:2013eea,Hod:2013nn}. The numerical results are presented in Section~\ref{sec6}, where we analyse some properties, in particular the non-uniqueness of charged BHs in the our model. Some further remarks are presented in Section~\ref{sec7}.

\section{The Einstein-Maxwell-scalar model}
\label{sec2}

\subsection{The action and the field equations}
We consider the Einstein-Maxwell-scalar model described by the following action
\begin{eqnarray}
\label{action}
 \mathcal{S}=\int d^4 x
\sqrt{-g}
\left[
\frac{R}{16\pi G}
-\frac{1}{4}F_{\alpha\beta}F^{
\alpha\beta}
-D_\alpha \Psi^*D^\alpha \Psi-U(\left|\Psi\right|)
\right]~,
\end{eqnarray}
where $G$ is the gravitational constant, $R$ is the Ricci scalar associated with the
spacetime metric $g_{\mu\nu}$, which has determinant $g$,
 $F_{\alpha\beta} =\partial_\alpha A_\beta - \partial_\beta A_\alpha$ is the Maxwell 2-form, $A_\alpha$ is the gauge 4-potential, $\Psi$ is a complex scalar field, `*' denotes complex conjugate  
and
\begin{eqnarray}
 D_\alpha\Psi\equiv \nabla_\alpha \Psi + iq A_\alpha \Psi \ , 
\end{eqnarray}
is the gauge covariant derivative, where $q$ is the gauge coupling constant and $\nabla$ is the geometric covariant derivative.
  $ U(|\Psi|)>0$ denotes the scalar potential, which in this work is taken to be always non-negative; the scalar  mass is defined by  $\mu^2\equiv  (d^2 U/d |\Psi|^2)\big|_{\Psi=0}$.

The Einstein--Maxwell-scalar field equations, 
obtained by varying (\ref{action}) with respect to the metric, scalar field and electromagnetic field, are, respectively,
\begin{eqnarray}
\label{E-eqs}
&&
R_{\alpha\beta}-\frac{1}{2}g_{\alpha\beta}R=8 \pi G \left[ T_{\alpha\beta} ^{\rm (EM)} +T_{\alpha\beta}^{(\Psi)} \right] \ , 
\\
\label{Ms-eqs} 
&&
D_{\alpha}D^{\alpha}\Psi=\frac{d U}{d\left|\Psi\right|^2} \Psi \ , \qquad 
~~\nabla_{\alpha}F^{\beta \alpha}=
iq \big [\Psi^*(D^\beta \Psi)-(D^{\beta}\Psi^*) \Psi )   \big ]~
\equiv q j^\beta ~,
\end{eqnarray}  
where  the two components of the energy-momentum tensor are
\begin{eqnarray}
\label{tmunu} 
T_{\alpha\beta}^{\rm (EM)} 
&=&
F_\alpha^{~\gamma}F_{\beta\gamma} - \frac{1}{4}g_{\alpha\beta}F_{\gamma\delta}F^{\gamma\delta} \ ,~
\\
\nonumber
T_{\alpha\beta}^{(\Psi)}
&=&
 D_\alpha\Psi^* D_\beta\Psi 
+D_\beta\Psi^* D_\alpha\Psi  
-g_{\alpha\beta}  \left[ \frac{1}{2} g^{\gamma\delta} 
 ( D_\gamma \Psi^* D_\delta\Psi+
D_\delta\Psi^* D_\gamma\Psi) +U(\left|\Psi\right|) \right]
 \ .
\end{eqnarray} 
This model is invariant under a local $U(1)$ gauge transformation 
\begin{eqnarray}
\label{gauge-transf}
\Psi \to \Psi e^{-i q \chi(x^\alpha)}\ , \qquad A_\beta\to A_\beta +\partial_\beta \chi(x^\alpha) \ ,
\end{eqnarray}
where $\chi(x^\alpha)$ is any real function.
The Maxwell equations in (\ref{Ms-eqs}) define the 4-current $j^\alpha$, which is conserved, $\nabla_\alpha j^\alpha=0$.

\subsection{The ansatz}
\label{secansatz}
For addressing spherically symmetric
 solutions we choose the following ansatz: for the metric,
\begin{eqnarray}
\label{metric}
ds^2=-N(r)\sigma^2(r)dt^2+\frac{dr^2}{N(r)}+r^2(d\theta^2+\sin^2\theta d\varphi^2)\ , \qquad {\rm with}~~
N(r)\equiv 1-\frac{2m(r)}{r} \ ,
\end{eqnarray}
where $t$ is a time coordinate (outside the horizon), $r$ is the areal radius and $\theta,\varphi$ are the standard spherical coordinates; for the scalar field and 4-potential,
\begin{eqnarray}
\Psi=\psi(r) e^{-i w t}\ , \qquad A= V(r) dt \ ,
\label{scaans}
\end{eqnarray}
where $w$ is the (real) oscillation frequency of the scalar field. The ansatz therefore introduces four radial functions: $\sigma(r),m(r),\psi(r),V(r)$.  The corresponding field equations, resulting from (\ref{E-eqs})-(\ref{Ms-eqs})
 read,\footnote{There is a also the constraint eq.
\begin{eqnarray}
\label{constr}
\frac{N''}{2}
+\left(\frac{\sigma'}{r}+\sigma''\right)\frac{N}{\sigma}
+\left(\frac{1}{r}+\frac{3\sigma'}{2\sigma}\right)N'
+8\pi G
\left[
 N\psi'^2
-\frac{V'^2}{2\sigma^2}
-\frac{(w-q V)^2\psi^2}{N\sigma^2}
+U(\psi)
\right]=0 \ ,
\end{eqnarray}
which is a combination of (\ref{eqm})-(\ref{eqpsi}).}
 denoting radial derivatives by ``primes",
\begin{eqnarray}
&&
\label{eqm}
m'=4\pi G r^2
\left[
\frac{V'^2}{2\sigma^2}+N\psi'^2+U(\psi)+\frac{(w-q V)^2}{N\sigma^2}\psi^2
\right] \ ,
\\
&&
\sigma'=8\pi G r \sigma
\left[
\psi'^2+\frac{(w-q V)^2\psi^2}{N^2 \sigma^2}
\right] \ ,
\\
&&
V''+\left(\frac{2}{r}-\frac{\sigma'}{\sigma}\right)V'+\frac{2q (w-q V)\psi^2}{N}=0\ ,
\label{Veq}
\\
&&
\label{eqpsi}
\psi''+
\left(
\frac{2}{r}+\frac{N'}{N}+\frac{\sigma'}{\sigma}
\right)\psi'
+\frac{ (w-q V)^2 \psi}{N^2\sigma^2}
-\frac{1}{2N}\frac{dU}{d\psi}=0~.
\end{eqnarray}
%
Inspection of these equations reveals a number of features.  Firstly, outside a BH horizon, wherein $N>0$, the metric functions 
$m(r)$
and
$\sigma(r)$
 are  increasing functions of $r$, whereas the electric gauge potential $V(r)$ is a strictly monotonic radial function. 
The latter conclusion becomes clearer after gauge fixing, $cf.$ Section~\ref{secgauge}, and rewriting (\ref{Veq}) in the form~(\ref{BM1}) below. Secondly, the model possesses the gauge freedom:
\begin{eqnarray}
\label{rgs}
w\to w+\zeta \quad  {\rm and} \quad V\to V+\frac{\zeta}{q} \ ,
\end{eqnarray}
where $\zeta$ is a constant that will be fixed below. Thirdly, there is also a  discrete symmetry
\begin{eqnarray}
\label{ds}
 V\to -V \quad  {\rm and} \quad q \to -q   \ ,
\end{eqnarray}
  which allows us to consider the case $q\geqslant 0$ only.

Within this framework,
the only non-vanishing component of the conserved 4-current is the temporal one:
\begin{eqnarray}
\label{jt}
j^t=\frac{2(q V-w)\psi^2}{N\sigma^2}\ ,
\end{eqnarray}
the associated Noether charge, $Q_N$, which is interpreted as the particle number, being
\begin{eqnarray}
\label{Noether}
Q_{N}=\frac{1}{4\pi}
\int d^3x \sqrt{-g} j^t =\int_{r_0}^{\infty}dr \frac{2r^2 (q V-w)\psi^2}{N\sigma }~,
\end{eqnarray}
where $r_0=r_h$ for BHs, with $r_h$ denoting the event horizon radius,  and $r_0=0$ for solitons.

For completeness, we include here the expression of the non-vanishing
components of the energy-momentum tensor
\begin{eqnarray}
\label{Tij}
&&
T_r^{r(EM)}=-\frac{V'^2}{2\sigma^2}\ , \qquad 
T_r^{r(\Psi)}=N\psi'^2+\frac{(w-q V)^2 \psi^2}{N\sigma^2}-U(\psi) \ ,
\\
\nonumber
&&
T_\theta^{\theta(EM)}=T_\varphi^{\varphi(EM)}=\frac{V'^2}{2\sigma^2}\ , \qquad
T_\theta^{\theta(\Psi)}=T_\varphi^{\varphi(\Psi)}=-N\psi'^2+\frac{(w-q V)^2\psi^2}{N\sigma^2}-U(\psi) \ ,
\\
\nonumber
&&
T_t^{t(EM)}=-\frac{V'^2}{2\sigma^2}\ , \qquad 
T_t^{t(\Psi)}=-N\psi'^2-\frac{(w-q V)^2\psi^2}{N\sigma^2}-U(\psi) \ .
\end{eqnarray}

\subsection{Virial identity}
A \textit{virial identity}, which is independent of the equations of motion, can be obtained for this model by using the approach in \cite{Heusler:1996ft}, which amounts to a scaling argument. Assuming a BH spacetime the result is:
\begin{eqnarray}
&&
\label{virial1}
\int_{r_h}^\infty dr~ r^2 \sigma
\left\{
     \left[
		             1-\frac{2r_h}{r}\left(1-\frac{m}{r}\right) 
		 \right]\psi'^2
+ \left(3-\frac{2r_h}{r} \right)U(\psi)
\right\}
\\
\nonumber
&&
=
\int_{r_h}^\infty dr~ r^2 
\left\{
\left(1-\frac{2r_h}{r}\right)\frac{V'^2}{2\sigma}
+
\left[
3-\frac{2r_h}{r}\left(1-\frac{3m}{r}\right)-\frac{8m}{r}
\right]\frac{(w-q V)^2 \psi^2}{N^2 \sigma}
\right\}~.
\end{eqnarray} 
On the one hand, both factors in front of
the (non-negative) scalar quantities on the $l.h.s.$  have a fixed, positive sign. Thus, the $l.h.s.$ integrand is non-negative, making the integral strictly positive. It immediately follows that for $V=0$ (no Maxwell field) and $w=0$, there can be no solution with a non-trivial scalar field. 
On the other hand, the factors in front of the (non-negative) Maxwell quantities
on the $r.h.s.$ of (\ref{virial1}) are indefinite, although they become positive asymptotically. Thus, for $V\neq 0$ and/or $w\neq 0$ a solution becomes possible (but not guaranteed). In fact, only $w\neq 0$ is insufficient to allow non-trivial solutions~\cite{Pena:1997cy}.

Another use of the relation (\ref{virial1}) is to check the 
accuracy of the numerical solutions. Indeed this was done in our work.

\section{Approximate solution and boundary conditions}
\label{apform}
In this work we are mostly interested in integrating the field equations (\ref{eqm})-(\ref{eqpsi}) to obtain BH solutions. Since this is done numerically, we should first discuss the asymptotic behaviours at the boundaries of the domain of integration. For spherically symmetric solutions this corresponds to the behaviour near the horizon and near spatial infinity.

\subsection{Near horizon expansion and the resonance condition}
Let the BH horizon be located at $r=r_h>0$. In this work we shall focus on non-extremal\footnote{The model is unlikely to possess
regular (on and outside a horizon) extremal BH solutions.
One hint in this direction is the absence of the usual 
 attractor solutions, $i.e.$ generalizations of the Bertotti-Robinson solution, with a metric $AdS_2\times S^2$.
A detailed investigation of extremal solutions will not, however, be addressed here.
}
 BHs, 
$i.e.$
$N(r)\sim (r-r_h)$ as $r\to r_h$.
Then, requiring finiteness of the energy-momentum
tensor (\ref{Tij}),
or of the current density (\ref{jt}), on the horizon, 
implies the following condition
\begin{eqnarray}
\psi(r_h) [w-q V(r_h)] =0 \ .
\end{eqnarray}
If one chooses $\psi(r_h)=0$, this turns out to imply that $d^k\psi/dr^{k}|_{r_h}=0$,
$i.e.$ the derivatives of the scalar field vanish order by order in a power series expansion
close to the horizon. This implies that the scalar field is trivial. 
Thus, in order to consider a non-trivial scalar field and finite physical quantities at the horizon, we are forced to consider the second choice
\begin{eqnarray}
\label{condition}
 w=q V(r_h)  \ .
\end{eqnarray}
We dub~(\ref{condition}) the \textit{resonance condition}. Choosing~(\ref{condition}), the scalar field can take a nonzero value at the horizon
and one can construct a power series expansion of the solution as $r\to r_h$. 
Without fixing the gauge freedom, 
this approximate solution reads
\begin{eqnarray}
&&
\nonumber
m(r)=\frac{r_h}{2}+4 \pi G r_h^2\left[\frac{v_1^2}{2\sigma_h^2}+U(\psi_h)\right](r-r_h)+\dots\  , \qquad 
\psi(r)=\psi_h-\frac{1}{2}
\frac{r_h \dot U(\psi_h)(r-r_h)}
{ 1-8\pi G r_h^2\left[\frac{v_1^2}{2\sigma_h^2}+U(\phi_h)\right]
}
+\dots
\\
&&
\sigma(r)=\sigma_h+\frac{8 \pi G \sigma_h\left[\frac{q^2 v_1^2 \psi_h^2}{\sigma_h^2}
+\frac{\dot U(\psi_h)^2}{4}\right]}{\left\{8\pi G r_h^2 \left[\frac{  v_1^2  }{2\sigma_h^2}+U(\psi_h)\right]-1\right\}^2} 
(r-r_h)+\dots \ , \qquad
V(r)=V(r_h)+v_1(r-r_h)+\dots \ ,
\end{eqnarray}
and contains 4 essential parameters,
\begin{eqnarray}
 \{
 \sigma_h,~\psi_h,~V(r_h),~ V'(r_h)
\} \ .
\end{eqnarray}
We have used the notation $\dot U(\psi_h)\equiv({dU}/{d\psi})|_{\psi_h}$.
%

\subsection{Far field expansion and the bound state condition}
We are interested in asymptotically flat solutions. Thus, at infinity, Minkowski spacetime is approached, 
while the scalar field and the gauge field strength vanish.
Without fixing the residual gauge freedom
(\ref{rgs}), one finds the approximate solution
\begin{eqnarray}
\label{inf}
&&
m(r)=M-\frac{4\pi G   Q_e^2}{2r }+\dots\ , \qquad 
\sigma(r)=1-\frac{4\pi G c_0^2 \mu^2}{\mu _\infty r}e^{-2 \mu_\infty r }\dots \ ,
\\
&&
\label{inf1}
V(r)=\Phi-\frac{Q_e}{r}+\dots \ , \qquad 
 \psi(r)=c_0 \frac{e^{-\mu_\infty  r}}{r}+\dots \ ,
\end{eqnarray}
where we have denoted
\begin{eqnarray}
 \mu_\infty \equiv \sqrt{\mu^2-(w-q \Phi)^2}~.
\end{eqnarray}
The free parameters in this expansion are
\begin{eqnarray}
 \{
M,~\Phi,~Q_e,~c_0
\} \ ,
\end{eqnarray}
where $M$ and $Q_e$ are the ADM mass and total electric charge,
  $\Phi$ is the asymptotic value of the electrostatic potential,
	while $c_0$ is an arbitrary constant.

From the above asymptotics, one notices the following \textit{bound state} condition
\begin{eqnarray}
\label{bsc}
w-q \Phi \leqslant \mu~.
\end{eqnarray}

\section{Gauge fixing, quantities of interest and scaling symmetries}
\label{sec4}

\subsection{Fixing the gauge}
\label{secgauge}
As discussed above, the model possesses  the residual gauge symmetry~(\ref{rgs}). In principle, the gauge choice is arbitrary.
However, not all gauge choices are physical for the problem at hand; for instance, they may not be compatible with the boundary conditions.
A discussion of these aspects for standard model solitons can be found in~\cite{Brihaye:1992jk}.
Two possible gauge choices are
\begin{equation}
V(r_h)=0 \ ,
\label{gaugec}
\end{equation}
 or
\begin{equation}
V(\infty)=0 \ .
\label{gaugec2}
\end{equation}
Our numerical results were found for the first choice.
Then the resonance condition~(\ref{condition})
implies $w=0$. It follows that the complex scalar field reduces to its amplitude, $cf.$~(\ref{scaans}). Consequently, this gauge choice is equivalent to consider the model~(\ref{action}) with a \textit{real}, rather than complex, scalar field.

After fixing the gauge in the way just described, the matter Lagrangian of the model 
can be written in the suggestive form
\begin{eqnarray}
\label{Ln}
\mathcal{L}_{\rm matter}=
-\frac{1}{4}F_{\alpha\beta}F^{\alpha\beta}
-\partial_\alpha \psi \partial^\alpha \psi-A_\alpha A^\alpha \psi^2-U(\psi) \ .
\end{eqnarray}
This can be interpreted as the scalar field endowing the gauge field with a dynamical mass term.
 In this gauge, the potential at infinity, $\Phi$,  is also the chemical potential, the difference between the values of the electric potential at infinity and  
at the horizon. Moroever,  the bound state condition (\ref{bsc}) implies $ \Phi\leqslant-\mu/q$.

We emphasize, however, that similar results are found for the second choice (\ref{gaugec2}).
Formally, passing from one gauge choice to another is provided by the  relation (\ref{rgs}),
with $\zeta=q \Phi $.
The physical results are, of course, independent of the gauge choice.

\subsection{Quantities of interest and measures of hairiness}
Let us now introduce some physical quantities of interest for the solutions we shall be discussing.
The Hawking temperature, $T_H$, 
and the event horizon area, $A_H$, of a solution are found from the horizon data,
\begin{eqnarray}
T_H=\frac{N'(r_h)\sigma(r_h)}{4\pi}\ , \qquad A_H=4\pi r_h^2 \ .
\end{eqnarray}
On the other hand, the ADM mass $M$,
the total electric charge $Q_e$
and the chemical potential $\Phi$
are determined by the far field asymptotics
(\ref{inf})-(\ref{inf1}).

For the chosen  gauge (\ref{gaugec}), $w=V(r_h)=0$, one finds the following intuitive decomposition of the total electric charge
\begin{eqnarray}
Q_e=Q_H+q Q_{N}\ , \qquad {\rm where} \qquad Q_H=\frac{1}{4\pi}\oint_H dS_{r}F^{tr}= 
\frac{r_h^2 v_1}{\sigma_h} \ ,
\end{eqnarray} 
where $Q_N $ is the Noether charge given by (\ref{Noether}),
while
$Q_H$ corresponds to the horizon electric charge.
Thus, the total electric charge is the sum of the electric charge within the horizon plus the Noether charge outside the horizon - which counts the  number of scalar particles - multiplied by the charge of a single particle. This decomposition suggests defining the following \textit{measure of hairiness}, denoted $h$:
\begin{eqnarray}
 h \equiv \frac{q Q_{N}}{Q_e}=1- \frac{Q_H}{Q_e}\ .
\label{ph}
\end{eqnarray} 
This measure takes the value $h=0$ for RN BH, which has no scalar field, $\psi=0$, and is 0\% hairy,
and takes the value $h=1$
for a soliton, for which $r_h=0$ and it is 100\% hairy.

Another possible measure of hairiness is  \cite{Delgado:2016zxv}
\begin{eqnarray}
p  \equiv \frac{M_H}{M}\ , ~
\end{eqnarray} 
where 
$M_H=\frac{1}{2}T_H A_H$
is the horizon mass. This corresponds to the fraction of the ADM mass which is stored inside the horizon. This measure, however, does not give $p=1$ for RN BHs, since part of the spacetime energy is due to the electromagnetic field outside the horizon. Thus, this measure is sharper for uncharged BHs, such as Kerr BHs with synchronised scalar~\cite{Herdeiro:2014goa} or Proca hair~\cite{Herdeiro:2016tmi}.

\subsection{Symmetries and scalings}
Below we shall focus on the  
simplest 
potential in the $Q$-ball literature~\cite{Coleman:1985ki}:
\begin{eqnarray}
\label{potential}
U(\psi)=\mu^2\psi^2-\lambda \psi^4+\nu \psi^6 \ .
\end{eqnarray}
As before, $\mu$ is the scalar field mass;  
$\lambda,\nu$ are positive parameters controlling the self-interactions of the scalar field.

Inspection of the field equations, with the choice (\ref{potential}), shows the existence of the  following 
scaling
symmetries of the (spherically ymmetric) model:\footnote{All functions or constants which are not explicitly mentioned
do not change under the corresponding transformation.}
\begin{eqnarray}
 &&
i)~~ t\to a t,~~V\to V/a,~~\sigma \to \sigma/a\ ,
\\
&& 
ii)~~ r\to a r,~~m\to a m,~~q\to q/a,~~\mu\to \mu/a,~~\lambda\to \lambda/a^2,~~ \nu\to \nu/a^2\ , 
\\
&& 
iii)~~ \phi\to a \phi,~~V\to  a V,~~q\to q/a,~~\lambda\to \lambda/a^2,~~ \nu\to \nu/a^4,~~ G\to G/a^2\ ,
\end{eqnarray}
where $a$ is an arbitrary non-zero parameter.
Symmetry $i)$ is fixed when imposing the boundary condition 
$\sigma(\infty)=1$.
As for the (ungauged) $Q$-balls case,  symmetries $ii)$ and $iii)$
are used to set $\mu=1$, $\lambda=1$
in the numerics.\footnote{$iii)$ can be used to set $G=1$, a choice employed for the (usual mini-)boson stars. 
However, the numerical study of the solutions starting with the probe limit is  
more intricate in this case. }
 Thus we take
\begin{eqnarray}
r\to r /\mu\ , \qquad m \to m/\mu\ , \qquad 
\phi \to \phi \mu/\sqrt{\lambda}\ , \qquad 
V\to V \mu/\sqrt{\lambda} \qquad
{\rm and} \qquad q\to q \sqrt{\lambda} \ .
\end{eqnarray}

At the end of this procedure, the model possesses
three independent dimensionless input parameters
\begin{eqnarray}
 &&
\alpha^2\equiv \frac{4\pi G \mu^2}{\lambda}\ , \qquad \beta^2\equiv \frac{\nu \mu^2}{\lambda^2}\ , \qquad e\equiv \frac{q}{\sqrt{\lambda}} \ .
\label{inputp}
\end{eqnarray}
For completeness, we exhibit here the
reduced Lagrangian of the rescaled model:
 \begin{eqnarray}
 &&
\mathcal{L}_{\rm eff}=\sigma \frac{dm}{dr}-\alpha^2
\sigma
\left[
N \left(\frac{d\psi}{dr}\right)^2 r^2
- \left(\frac{dV}{dr}\right)^2  \frac{r^2}{2\sigma^2}+
(\psi^2-\psi^4+\beta^2 \psi^6)r^2
-\frac{e^2 V^2 r^2 \psi^2}{N\sigma^2}
\right] \ .
\end{eqnarray}

Working with scaled variables,
the first law of thermodynamics in this model reads
\begin{eqnarray}
\label{1st}
dM=\frac{1}{4}T_H dA_H + \alpha^2 \Phi dQ_e \ ,
\end{eqnarray}
while
the Smarr law is
\begin{eqnarray}
\label{Smarr}
M=\frac{1}{2}T_H A_H+ \alpha^2 \left[M_{(\psi)} + \Phi Q_e\right] \ ,
\end{eqnarray}
where $M_{(\psi)}$ is the mass outside the horizon stored in the scalar field
\begin{eqnarray}
\label{Mint}
 M_{(\psi)}= 2 \int_{r_h}^\infty
dr~r^2\sigma
\left[
\frac{e^2 V^2 \psi^2}{N \sigma^2}-U(\psi)
\right] \ .
\end{eqnarray}

\section{Circumventing  no-hair theorems}
\label{sec5}
Different results in the literature establish the impossibility of having BHs with scalar hair in the model (\ref{action}), or endowing RN BHs with minimially coupled scalar hair. Let us clarify how these results are actually circumvented by our setup.

\subsection{The Mayo-Bekenstein no-go result}
One of the best known no-hair theorems for BHs that applies to our model~(\ref{action}) was establish 
by Mayo and Bekenstein~\cite{Mayo:1996mv}. The 
no-hair result is based on the following arguments, herein adapted to our framework (see also the discussion in~\cite{Hong:2019mcj}).
One starts with the equation for the electric potential~(\ref{Veq}), which after the gauge fixing (\ref{gaugec}) is
written in the following form
\begin{eqnarray}
\label{BM1}
\left(\frac{r^2 V'}{\sigma}\right)'= \frac{2q^2 r^2 V \psi^2}{N \sigma}~.
\end{eqnarray}
Multiplying (\ref{BM1}) by $V(r)$, rearranging the expression and integrating both sides from $r_h$ to infinity yields
\begin{eqnarray}
\label{BM2}
\left(\frac{r^2 V V'}{\sigma}\right)\Big|_{r=\infty}=Q_e\Phi= 
\int_{r_h}^\infty\left[\frac{r^2 V'^2}{\sigma}+
\frac{2q^2 r^2 V^2 \psi^2}{N \sigma}\right]dr \ ,
\end{eqnarray}
using the gauge choice $V(r_h)=0$ and the asymptotic behaviours (\ref{inf})-(\ref{inf1}). The $r.h.s.$
is strictly positive; thus, solutions necessarily have $\Phi\neq 0$.

If $\Phi\neq0$, then this chemical potential  provides an effective $tachyonic$ mass for $\psi$,
	$\mu_{\rm tac}^2=-q^2 \Phi^2$, as can be seen from (\ref{Ln}). 
	If, \textit{as assumed} by Mayo and Bekenstein, there is no mass term coming from the potential $U(\psi)$, 
	then this tachyonic mass term is not compatible with a proper asymptoptic behaviour of the scalar field. Therefore, one needs to impose $\Phi=0$. Then, from (\ref{BM2})
one concludes that the electric potential vanishes everywhere, reducing the problem to that of uncharged BHs.
 Other no-hair theorems, or for instance the virial identity~(\ref{virial1}), then imply that the scalar field must also vanish.

It should now be clear that the reasoning in~\cite{Mayo:1996mv} holds for a massless scalar field only, $i.e.$ $\mu=0$, which will not be the case here.

\subsection{The Pe\~na-Sudarsly no-go result}

The  no-hair theorem due to Pe\~na and Sudarsky~\cite{Pena:1997cy}
holds for Einstein's gravity minimally coupled to generic matter fields
fulfilling the the weak energy condition,  which holds for the model in this work, 
together 
with the condition
\begin{eqnarray}
\label{condPS}
 T_\theta^\theta\leqslant T_r^r \ .
\end{eqnarray}
One can check that (\ref{condPS})
is satisfied by the scalar components of the total energy-momentum tensor,
$
 T_\theta^{\theta(\Psi)}- T_r^{r(\Psi)}=-2N\psi'^2<0.
$
Thus,  for a  pure Einstein-Klein-Gordon model, a static, spherically symmetric BH cannot support scalar hair \cite{Pena:1997cy}.
However, the relation (\ref{condPS})
fails to be satisfied for the vector part of the total energy-momentum tensor,
$
 T_\theta^{\theta(EM)}- T_r^{r(EM)}=V'^2/\sigma^2>0,
$
and thus non-Schwarzschild BHs are possible in Einstein-Maxwell theory; indeed this is the RN solution.
Similarly, the full energy-momentum tensor (\ref{Tij})
does not satisfy the relation (\ref{condPS}),
and thus one cannot use this argument to exclude the existence of 
charged BH  solutions with scalar hair in our model with the potential~(\ref{potential}).

\subsection{The Hod no-charged-BH bomb result}
For a given BH solution, the existence of an instability zero mode, for any type of perturbation, is a smoking gun for a new family of solutions. There are many examples of this pattern. In the context of our work the most relevant example is the bifuraction of the Kerr family into BHs with synchronised scalar~\cite{Herdeiro:2014goa} or Proca~\cite{Herdeiro:2016tmi} hair, due to the superradiant instability of a massive scalar or vector field. In this case, the zero mode of the superradiant instability of a massive bosonic field is a mode with frequency $w$ and azimuthal harmonic index $m$ which obeys the synchronisation condition
\begin{equation}
w=m\Omega_H \ ,
\end{equation}
where $\Omega_H$ is the angular velocity of the Kerr BH horizon.

The RN BH is afflicted by charged superradiance~\cite{Bekenstein:1973mi}. The zero mode of the charged superradiant instability of a \textit{charged} bosonic field is a mode with frequency $w$ and charge $q$ obeying the resonance condition~(\ref{condition}). So, in principle, the RN family could bifurcate towards a new family of BH solutions with massive bosonic hair. It turns out, however, that there are no zero modes of the instability that are \textit{also} bound states.\footnote{There are, however, zero modes that are \textit{marginally} bound states~\cite{Degollado:2013eqa,Sampaio:2014swa}.} This was shown by Hod in the case of a test, massive, charged scalar field on the RN background~\cite{Hod:2013eea,Hod:2013nn}, and, in principle, rules out the existence of RN BHs with scalar hair that are asymptotically flat. However, the analysis of Hod is based on a linear Klein-Gordon equation, without self-interactions. As we shall see, this is compatible with our results, since not only the self-interactions are fundamental, but also, the hairy BHs do not bifurcate from a zero mode around the RN background.

\subsection{Other Bekenstein-type relations}
\label{secbek}
It is of interest to derive other general relations, which provide further insight into the existence of the hairy BH solutions we will be reporting.

First, for the gauge choice employed here (\ref{gaugec}),
the scalar field equation can be written in the suggestive form
\begin{eqnarray} 
\label{eqp1}
(r^2 N \sigma \psi')'=  r^2\sigma \mu_{\rm eff}^2 \psi~,
\end{eqnarray} 
where we have defined the \textit{effective mass}
\begin{eqnarray}
\mu_{\rm eff}^2\equiv \mu^2-\frac{q^2 V^2}{N\sigma^2}-2\lambda \psi^4+3\nu \psi^6~.
\label{mueff}
\end{eqnarray}
We observe that $\mu_{\rm eff}^2>0$ asymptotically, but actually $\mu_{\rm eff}^2$ must change sign. Indeed, integrating (\ref{eqp1}) between the horizon and infinity
one finds  
\begin{eqnarray}
\label{cond12}
 \int _{r_h}^\infty dr r^2\sigma \mu_{\rm eff}^2 \psi=0 \ ,
\end{eqnarray} 
since $N(r_h)=0$ and $\psi'(\infty)=0$
For a nodeless scalar field (the case in this work),
this implies that $\mu_{\rm eff}^2$ takes both positive and
negative values outside the horizon.

Second, multiplying (\ref{eqp1}) by either $\psi$ or $ \mu_{\rm eff}^2$ and integrating between the horizon and infinity, one finds the following Bekenstein-type identities
\begin{eqnarray} 
&&
\label{Bi1}
r^2 N \sigma \psi \psi' \bigg|_{r_h}^\infty= 
\int _{r_h}^\infty dr
\left(
r^2 N \sigma \psi'^2 +
 r^2\sigma \mu_{\rm eff}^2 \psi^2  
\right) \ ,
\\
\label{Bi2}
&&
r^2 N \sigma \psi' \mu_{\rm eff}^2 \bigg|_{r_h}^\infty= 
\int _{r_h}^\infty dr
\left[
r^2 N \sigma \psi'  \frac{d \mu_{\rm eff}^2}{dr }+
\frac{1}{2}r^2\sigma  ( \mu_{\rm eff}^2 )^2
\right]  \ .
\end{eqnarray}

Since $N(r_h)=0$ and $\psi(\infty)=\psi'(\infty)=0$ the $l.h.s.$ of (\ref{Bi1})-(\ref{Bi2}) vanishes. Then, the condition for the existence of physical solutions is that the one term in the $r.h.s.$ integrand which is not manifestly positive is actually negative in some radial interval; that is, 
\begin{eqnarray}
\label{cond1}
\mu_{\rm eff}^2<0\ , \qquad {\rm and} \qquad \psi'  \frac{d \mu_{\rm eff}^2}{dr }<0 \ ,
\end{eqnarray} 
for some range of $r>r_h$. In particular, this confirms that the effective mass square must change sign.
%

%

\section{Numerical results}
\label{sec6}

\subsection{The numerical approach}
%
The set of four ordinary differential equations
(\ref{eqm})-(\ref{eqpsi})
 is solved under suitable boundary conditions
which result from the near horizon/far field approximate form of the solution, as described in Section~\ref{apform} for BHs and also below for solitons.
In most of the numerics we have employed a collocation method for boundary value
ordinary differential equations equipped with an adaptive mesh selection procedure 
\cite{colsys}. 
Typical mesh sizes include $10^{3}-10^{4}$ points,
the relative accuracy of the solutions being
around $10^{-10}$.
A large part of the solutions were also constructed by using a standard Runge-Kutta ordinary differential
equation solver. In this approach we evaluate the initial conditions at $r=r_h+10^{-8} $, for global
tolerance $10^{-14}$, adjusting for the shooting parameters
$\psi(r_h)$ and $\sigma_h$
 and integrating towards $r\to \infty$.

\subsection{Solitonic solutions}

The model~(\ref{action}) possesses smooth particle-like solitonic solutions, which 
are recovered as the limit $r_h\to 0$
of the hairy BHs to be described below. 
These \textit{gauged boson stars} have been discussed in the literature, albeit less so than the usual uncharged boson stars.
Their study can be traced back, at least, to~\cite{Jetzer:1993nk,Jetzer:1989av}
which considered
gauged boson stars without a sextic term in the potential and $\lambda<0$, although the presence  of the quartic term is not crucial.
More recently, their were considered in~\cite{Pugliese:2013gsa}.  
We are not aware of any systematic discussion of the gauged boson stars with the $Q$-ball potential (\ref{potential}), \textit{a.k.a.} gravitating gauged $Q$-balls, in the literature; some partial results can be found in  
\cite{Prikas:2002ij,Brihaye:2014gua}.

The far field asymptotics of these globally regular configurations   
is similar to the BH case described in Section~\ref{apform}. Their small-$r$ form, however, is different and reads\footnote{We assume the gauge (\ref{gaugec}), thus $w=0$ and work with unscaled variables and a generic scalar field potential.}
\begin{eqnarray}
&&
m(r)=\frac{4\pi G}{3}
\left[
\frac{q^2 V_0^2 \psi_0^2}{\sigma_0^2}+U(\psi_0)
\right] r^3+\dots \ , \qquad \sigma(r)=\sigma_0+\frac{4\pi G q^2 V_0^2 \psi_0^2}{\sigma_0}r+\dots \ ,
\\
\nonumber
&&
\psi(r)=\psi_0+\frac{1}{6}
\left[
-\frac{q^2 V_0^2 }{\sigma_0^2}\psi_0+\frac{1}{2}\frac{dU(\psi)}{d\psi}\bigg|_{\psi_0}
\right]r^2+\dots\ , \qquad 
V(r)=V_0+\frac{1}{3}q^2V_0\psi_0^2 r^2+\dots \ ,
\end{eqnarray}
which contains three free parameters
$\sigma_0,V_0$ and $\psi_0$.
Note that the electric potential {\it does not vanish}
at $r=0$.
The first law in this case reads
\begin{eqnarray}
dM=\alpha^2 \Phi dQ_e~.
\end{eqnarray}

Whereas we shall not scan the domain of existence of the solitonic solutions, let us provide a sketchy overview of it. 
Firstly, fixing the input parameters of the model (\ref{inputp}) $(\alpha,\beta,e)$,
the gauged boson stars exist for a finite range of the 
chemical potential $\Phi$,
the upper limit being
fixed by the bound state condition (\ref{bsc}), 
$\Phi\leqslant -\mu/q$.
Secondly,  for given $(\alpha,\beta)$ and fixed  $\Phi$,
the solutions exist for a finite range of the gauge coupling constant
$0\leqslant q< q_{\rm max}$. This domain has been studied for non-self-interacting scalar fields~\cite{Jetzer:1993nk,Pugliese:2013gsa}, but not in the general case. 
Thirdly, in the uncharged limit, these solutions reduce to the standard ungauged boson stars. Formally, this limit is 
$q\to \epsilon$,
$V\to -w/\epsilon$ and $\epsilon \to 0$.\footnote{Alternatively, this limit is found
by  redefining $A_\alpha\to A_\alpha/q$
in the initial action (\ref{action}),
such that 
the Maxwell Lagrangian becomes $F_{\alpha\beta}F^{\alpha\beta}/q^2$,
while the covariant gauge derivative is
$ D_\alpha\Psi=\partial_\alpha \Psi + i  A_\alpha \Psi$.
Then $A_\alpha\to 0$ as $q\to 0$.
} 
Thus, for small $q$ the 
solutions are rather similar to  
the corresponding ungauged configurations,
while the electric potential
$V(r)$ is large and almost constant, with $V(0)\simeq \Phi$. Finally, for the full $Q$-ball potential (\ref{potential}) the solutions admit a flat spacetime limit wherein they become non-self-gravitating solitons, corresponding to $\alpha\rightarrow 0$. For this potential, 
	the solutions exist for a finite range of values of 
	$\alpha$. Again, this domain has not been discussed in the literature.

\subsection{The hairy BHs - test field limit}
 Before discussing the BH solutions of the full Einstein-Maxwell-gauged scalar system,
it is of interest to consider a decoupling limit of the model, which
 corresponds to ignoring 
the backreaction of the scalar (and possibly also the Maxwell) field(s) on a fixed BH background geometry.
The corresponding equations and  boundary behaviours 
result  directly from the general case discussed above.

The study  of the test field limit is technically easier and often instructive about the fully non-linear system. In this approximation, we shall consider two cases of interest.

\subsubsection{ Gauged $Q$-clouds on a Schwarzschild BH background}
First we consider the Maxwell-gauged scalar field system on a fixed \textit{Schwarzschild} background. This limit is obtained by taking $\alpha \to 0$ 
in the field equations, but keeping $r_h\neq 0$. 
Then the Einstein and matter fields equations
decouple. The
Schwarzschild background has $m(r)=r_h/2$ ($N(r)=1-r_h/r$) and $\sigma(r)=1$
in the metric ansatz (\ref{metric}).

In this decoupling limit, the input (physical) parameters to find solutions are $\{r_h,\beta,e\}$.
 The total mass-energy of the field configuration is computed as the integral
\begin{eqnarray}
E=\frac{1}{4\pi}\int d^3x \sqrt{-g} 
(
T_\alpha^\alpha-2T_t^t
)= M_{(\psi)} + \Phi Q_e\ .
\end{eqnarray}
The electric charge and chemical potential 
are computed from the far field asymptotics of 
the Maxwell field (\ref{inf1}), whereas $M_{(\psi)}$ is given by (\ref{Mint}).

Within this setup we have obtained non-trivial field configurations, dubbed \textit{gauged $Q$-clouds}, on a Schwarzschild background. It may seem surprising that such solutions exist on a Schwarzschild (rather than RN) background. But recall that the gauge condition (\ref{gaugec}) is fulfilled by the Schwarzschild horizon. This is somewhat similar to the fact that, as a test field, Maxwell's equations admit a spherically symmetric solution on the Schwarzschild background - see Section 2.1 in~\cite{Herdeiro:2015waa}. This is a linear version (on the Maxwell field) of the RN BH. Similarly, the gauged $Q$-clouds we are describing correspond to a decoupling limit (rather than a linearisation) of a charged BH with gauged scalar hair.\footnote{A similar situation is found 
$e.g.$
in Einstein-Yang-Mills-$SU(2)$ theory.
The Yang-Mills equations posess an exact solution on a Schwarzschild background
\cite{BoutalebJoutei:1979va},
which captures the 
basic features of the self-gravitating configurations
\cite{89}.
}

 {\small \hspace*{3.cm}{\it  } }
\begin{figure}[h!]
\hbox to\linewidth{\hss%
	\resizebox{9cm}{7cm}{\includegraphics{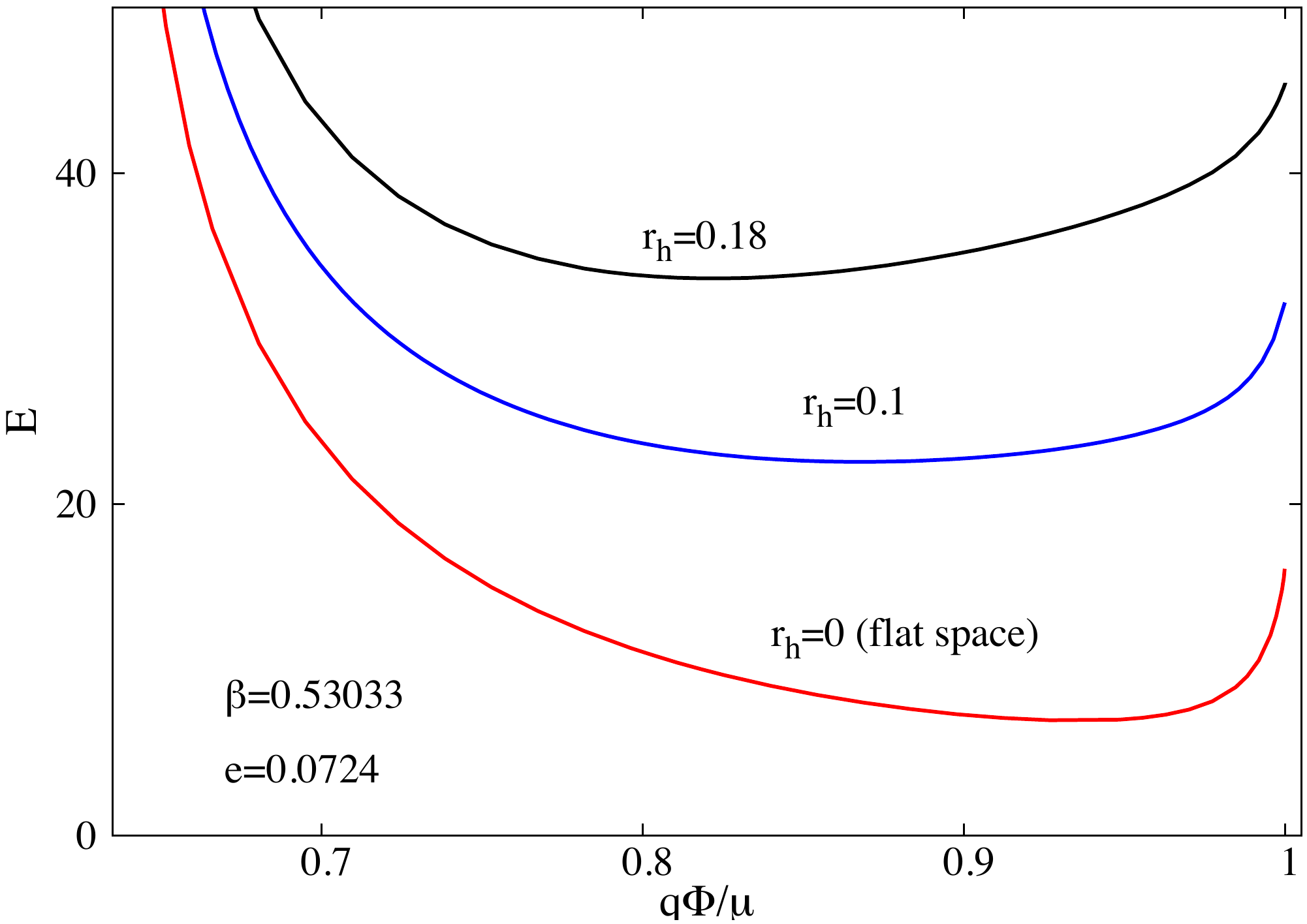}} 
	\resizebox{9cm}{7cm}{\includegraphics{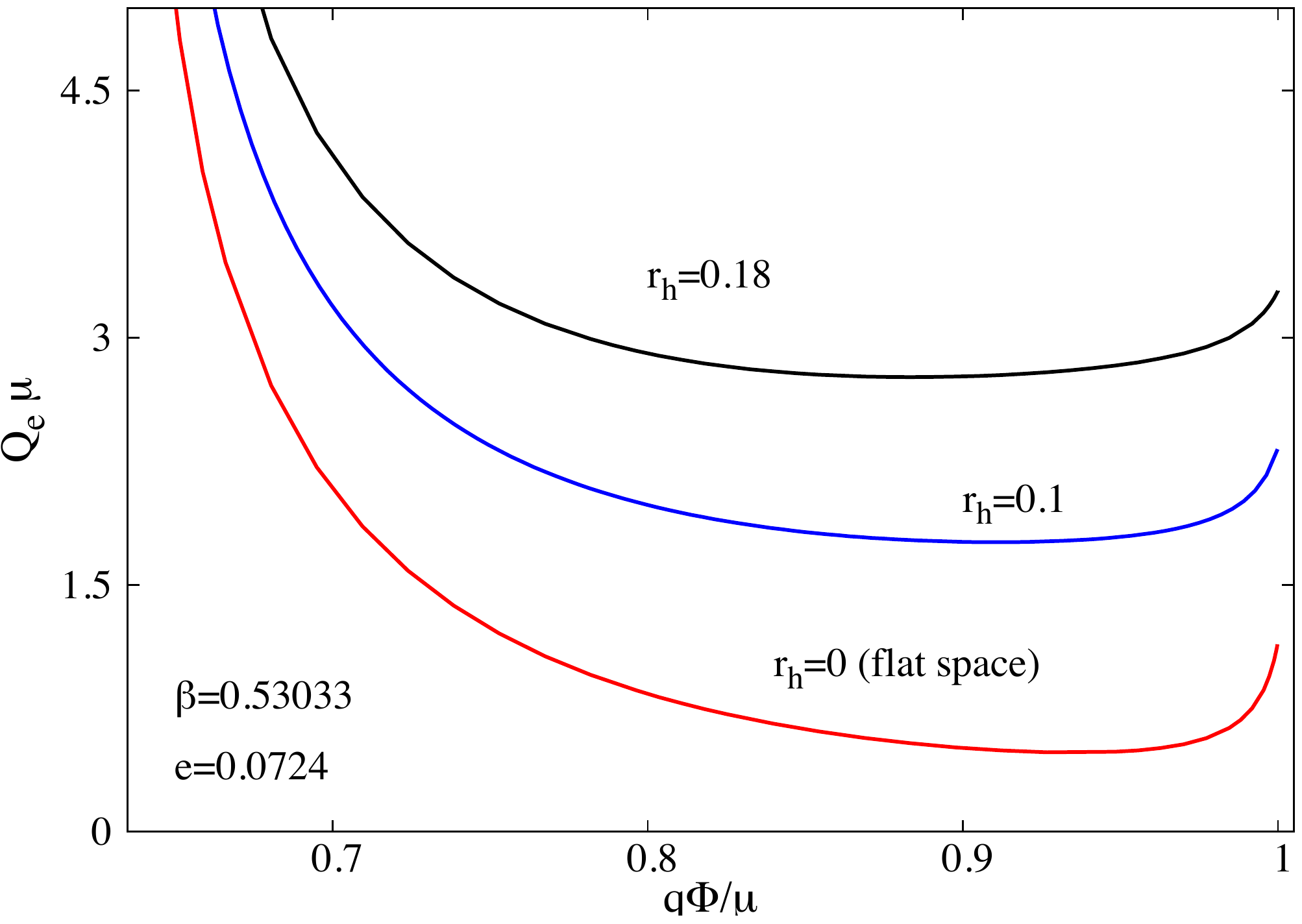}}  
\hss}
\caption{\small 
Gauged Q-clouds on fixed Schwarzschild BH backgrounds. Mass-energy (left panel) and electric charge (right panel) $vs.$ the chemical potential.  $r_h$ is the event horizon radius; $r_h=0$ the Minkowski spacetime limit. All quantities are given  in units set by $\mu$. 
}
\label{probe-Schw}
\end{figure}

Some salient features of these gauged $Q$-clouds are the following. Firstly, there is a mass gap: the solutions cannot have an arbitrarily small energy or electric charge and thus do not emerge as zero modes. This gap decreases with $r_h$ but is non-vanishing even in the flat spacetime limit $r_h=0$ - Fig.~\ref{probe-Schw} (left and right panels). As a related feature,  the scalar field never vanishes. 
Secondly, given a (Schwarzschild) BH background,
the solutions exist for some finite range of the (modulus of the) chemical potential
\begin{eqnarray}
\label{cond}
 \Phi_{\rm min}<|\Phi|<\mu/q,
\end{eqnarray}
a behavior similar to the flat spacetime case. This is also manifest in both panels of Fig.~\ref{probe-Schw}. 
%
Thirdly, for given $r_h$, the solutions exist for a finite range of the parameters 
$q,\beta$.

Let us emphasise that the non-linearities are key for the existence of these $Q$-clouds. 
Indeed, we can show that in the absence of scalar self-interaction no non-trivial solutions exist on the Schwarzschild background. To see this, observe that in the absence of self-interactions the effective mass (\ref{mueff}) reduces to 
\begin{eqnarray}
\mu_{\rm eff}^2=\mu^2-\frac{q^2 V^2}{N\sigma^2} \ .
\end{eqnarray}
It is clear that the effective mass is positive at both the horizon and asymptotically:  $\mu_{\rm eff}^2(r_h)=\mu^2>0$
and $\mu_{\rm eff}^2(\infty)=\mu^2-q^2 \Phi^2>0$. Recall the observation that in Section \ref{secbek} that solutions with the considered asymptotic behaviours require that $\mu_{\rm eff}^2$ must be negative in some range. 
Thus, there exists $r^*>r_h$
such that
\begin{eqnarray}
\label{cond12n}
\mu_{\rm eff}^2(r^*)<0 \ , \qquad  \frac{d\mu_{\rm 
eff}^2}{dr}\bigg|_{r=r^*}=0 \ , \qquad  \frac{d^2\mu_{\rm eff}^2}{dr^2}\bigg|_{r=r^*}>0~.
\end{eqnarray}
A straightfoward computation leads to the following relation, using the middle eq. in (\ref{cond12n}) and the field equation for the electric potential:
\begin{eqnarray}
\label{cond12n2}
\frac{d^2\mu_{\rm eff}^2}{dr^2}\bigg|_{r=r^*}=
-\frac{2q^2 r (2q^2 r V^2\psi^2+(r-r_h)V'^2)}{(r-r_h)^2}\bigg|_{r=r^*}<0 \ .
\end{eqnarray}
This contradicts the last condition in (\ref{cond12n}) 
(we recall that $r^*-r_h>0$).
Thus we conclude that the Schwarzschild BH does not support a gauged scalar cloud for the simple model
with a mass term only.
 
Finally, let us mention that the 
virial relation~(\ref{virial1})
takes a particularly simple form in the probe limit. It reads
\begin{eqnarray}
\label{virial12}
\int_{r_h}^\infty dr~ r^2  
\left\{
      N^2\psi'^2
+ \left(3-\frac{2r_h}{r} \right)U(\psi)
\right\}
=
\int_{r_h}^\infty dr~ r^2 
\left\{
\left(1-\frac{2r_h}{r}\right)\frac{V'^2}{2}
+
3e^2 V^2 \psi^2
\right\}~.
 \end{eqnarray}

\subsubsection{ Gauged $Q$-clouds on a RN background}
Now we consider a gauged scalar test field on a \textit{RN} background.
This limit is found by taking again the limit $\alpha \to 0$, but now rescaling also $V\to V/\alpha$ and $e\to \alpha e$.
The Einstein-Maxwell
and gauged scalar field sectors of the model decouple.  We are left solutions of eq. (\ref{eqpsi})
on a fixed RN BH background, which has 
 \begin{eqnarray}
N(r)=1-\frac{2M}{r}+\frac{Q^2}{r^2}\ , \qquad \sigma(r)=1 \ , \qquad V(r)=\frac{Q}{r_h}\left(1-\frac{r_h}{r}\right) \ , 
\end{eqnarray} 
in the metric ansatz (\ref{metric}) and gauge field ansatz (\ref{scaans}). $r_h$ is the largest root of $N(r_h)=0$.

This limit of the model has been considered in 
the recent work
\cite{Hong:2019mcj}.
We have confirmed independently the existence of the solutions
reported therein.
Our study of the backreacting solutions in the next section, however, suggests this limit is never approached by the
fully non-linear system, at least for the range of the parameters
explored so far.  


\subsection{The hairy BHs - non-perturbative results}
\begin{figure}[h!]
\hbox to\linewidth{\hss%
	\resizebox{9cm}{7cm}{\includegraphics{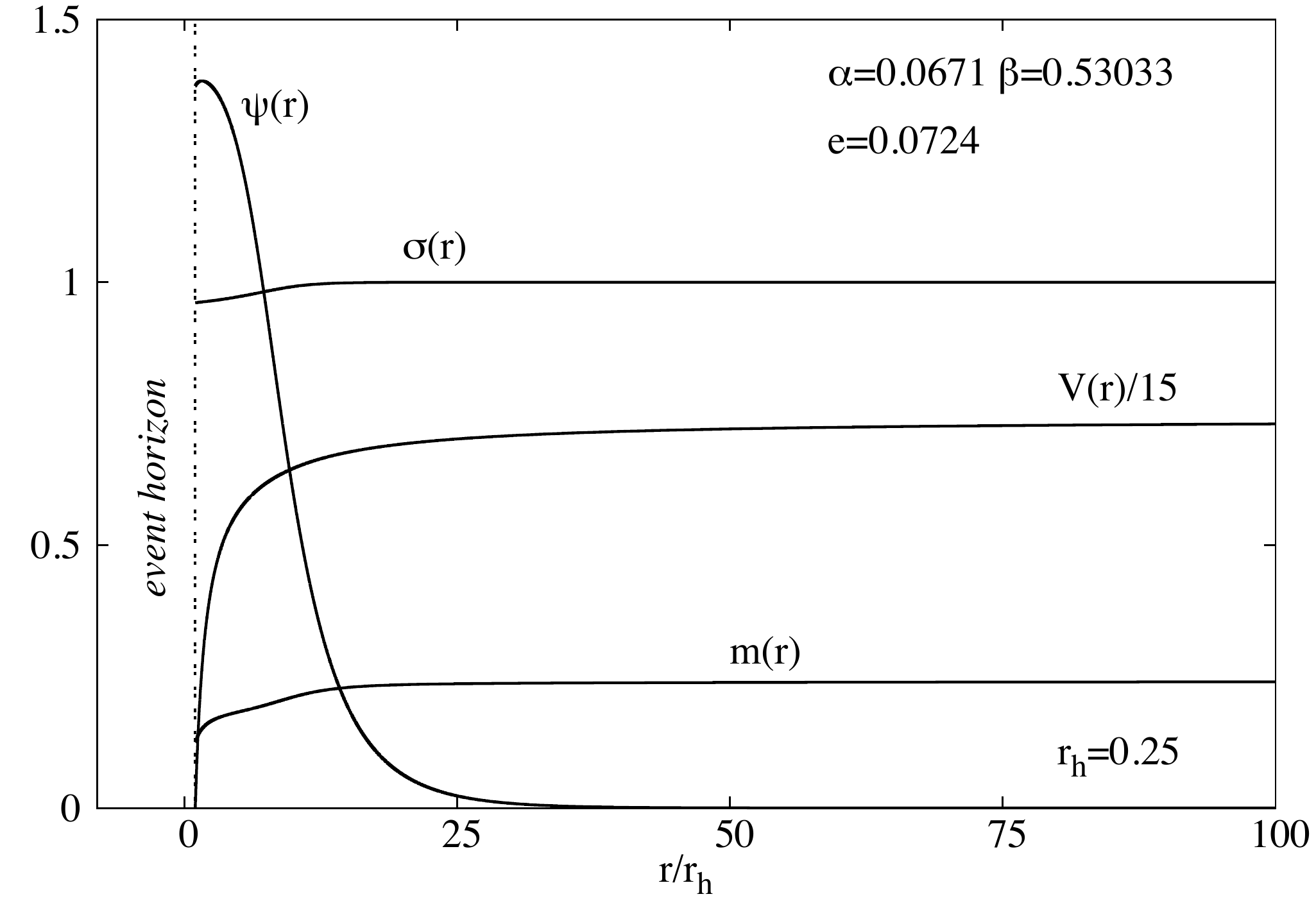}}   
\hss}
\caption{\small 
Profile functions  of an illustrative BH with gauged scalar hair, as functions of the radial coordinate. 
}
\label{profile}
\end{figure}

Let us now consider the fully non-linear solutions obtained by solving the equations of motion (\ref{eqm})-(\ref{eqpsi}) with the potential (\ref{potential}). We have explored these equations for a large set of physical parameters, 
$(\alpha,\beta,e)$.
The profile of a typical solution 
is shown in Fig.~\ref{profile}. As anticipated in Section~\ref{secansatz}, the functions $m(r),\sigma(r)$ are increasing radial functions and $V(r)$ is monotonic - in this case decreasing. Also, the scalar profile is nodeless and vanishes asymptotically.

We shall not pursue a complete scanning of the domain of existence of these solutions. 
Rather, we shall describe some sequences of solutions which, hopefully, are illustrative of general patterns.

 {\small \hspace*{3.cm}{\it  } }

\begin{figure}[t!]
\hbox to\linewidth{\hss%
	\resizebox{9cm}{7cm}{\includegraphics{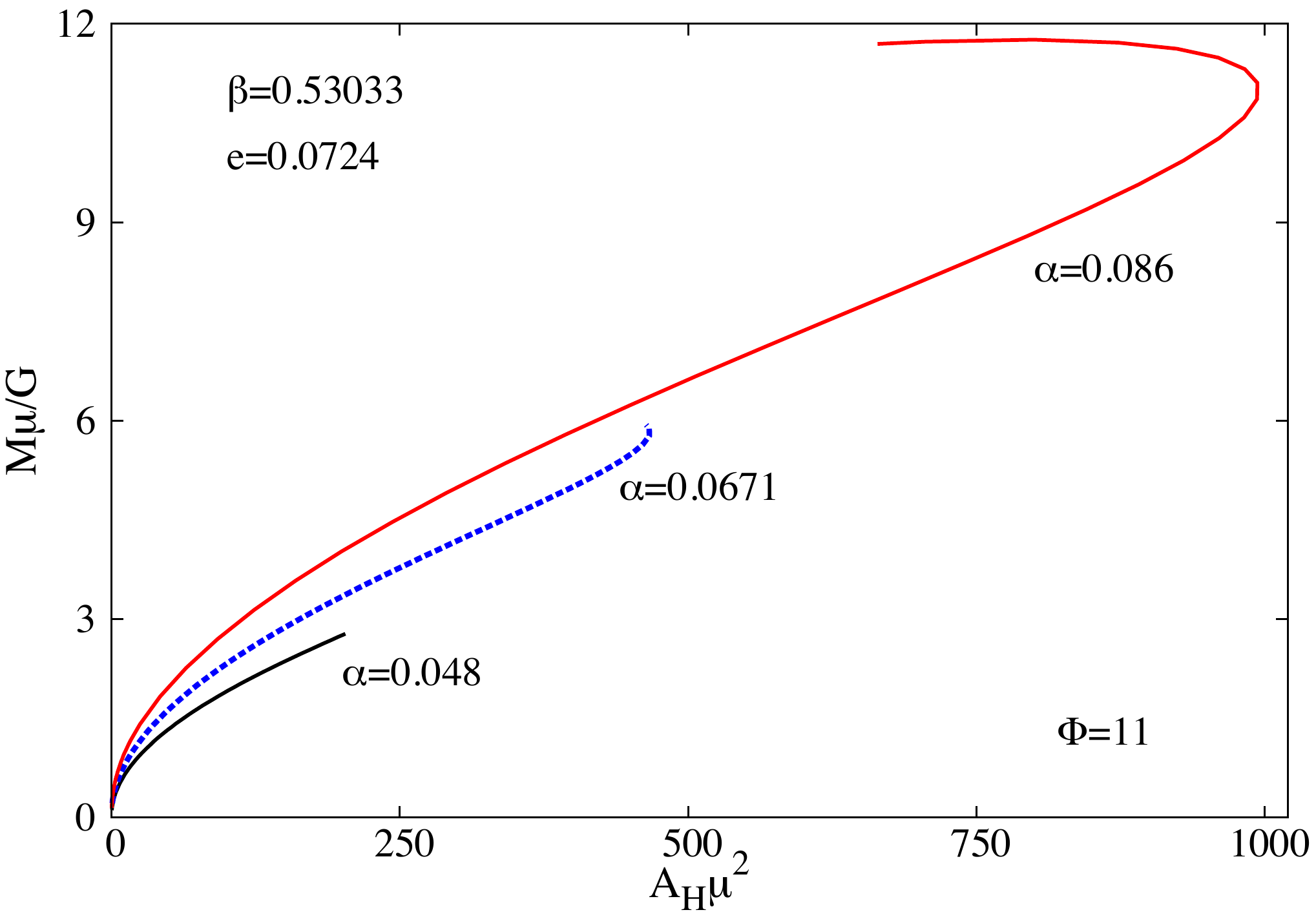}} 
	\resizebox{9cm}{7cm}{\includegraphics{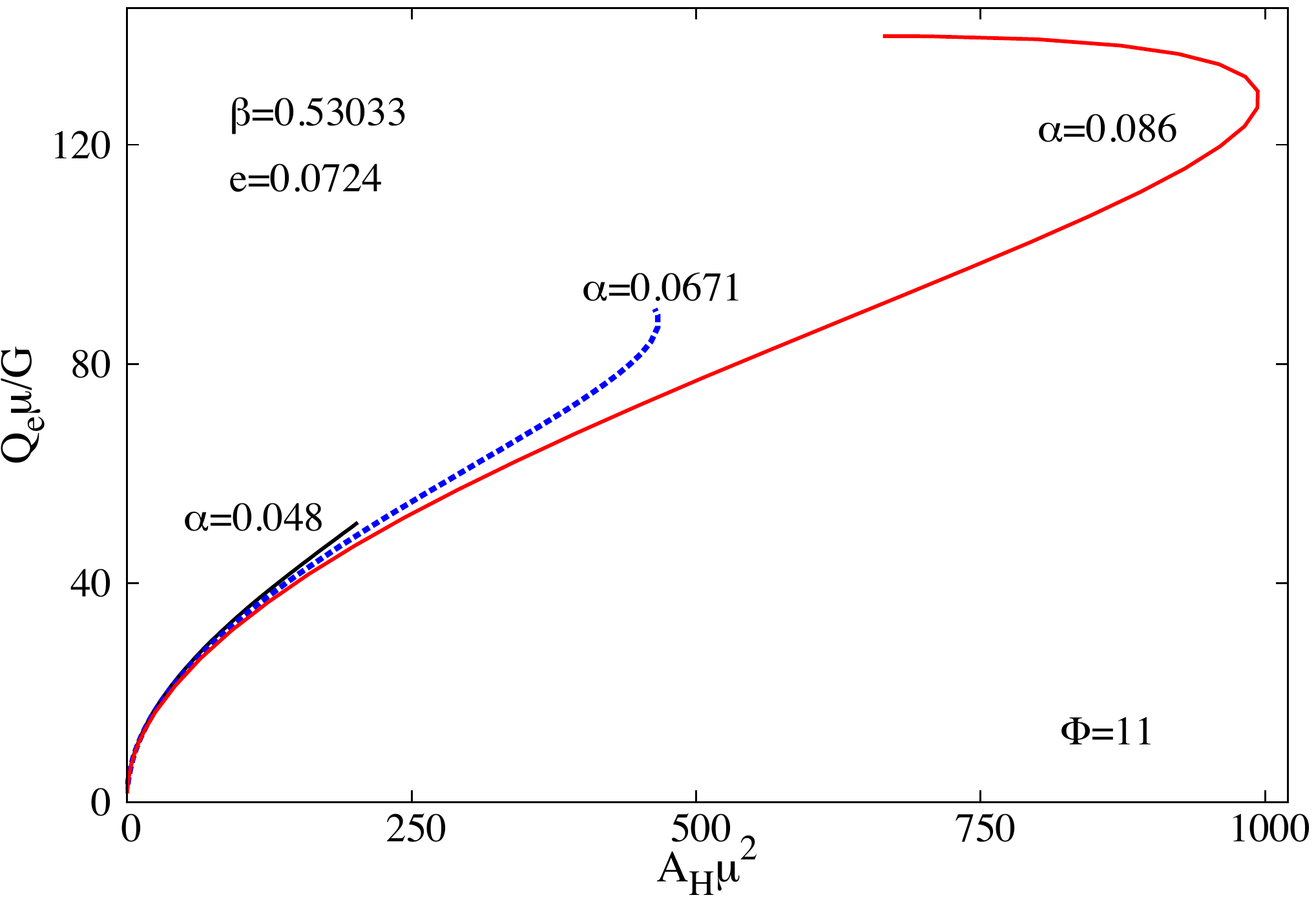}}  
\hss}
\caption{\small 
ADM mass $M$ (left panel) 
and electric charge
$Q_e$ of the RN BHs with resonant scalar hair, for a  fixed
chemical potential $\Phi$ and several values of the coupling constant
$\alpha$, $vs.$ the horizon area $A_H$.
}
\label{gravity1}
\end{figure}

\begin{figure}[t!]
\hbox to\linewidth{\hss%
	\resizebox{9cm}{7cm}{\includegraphics{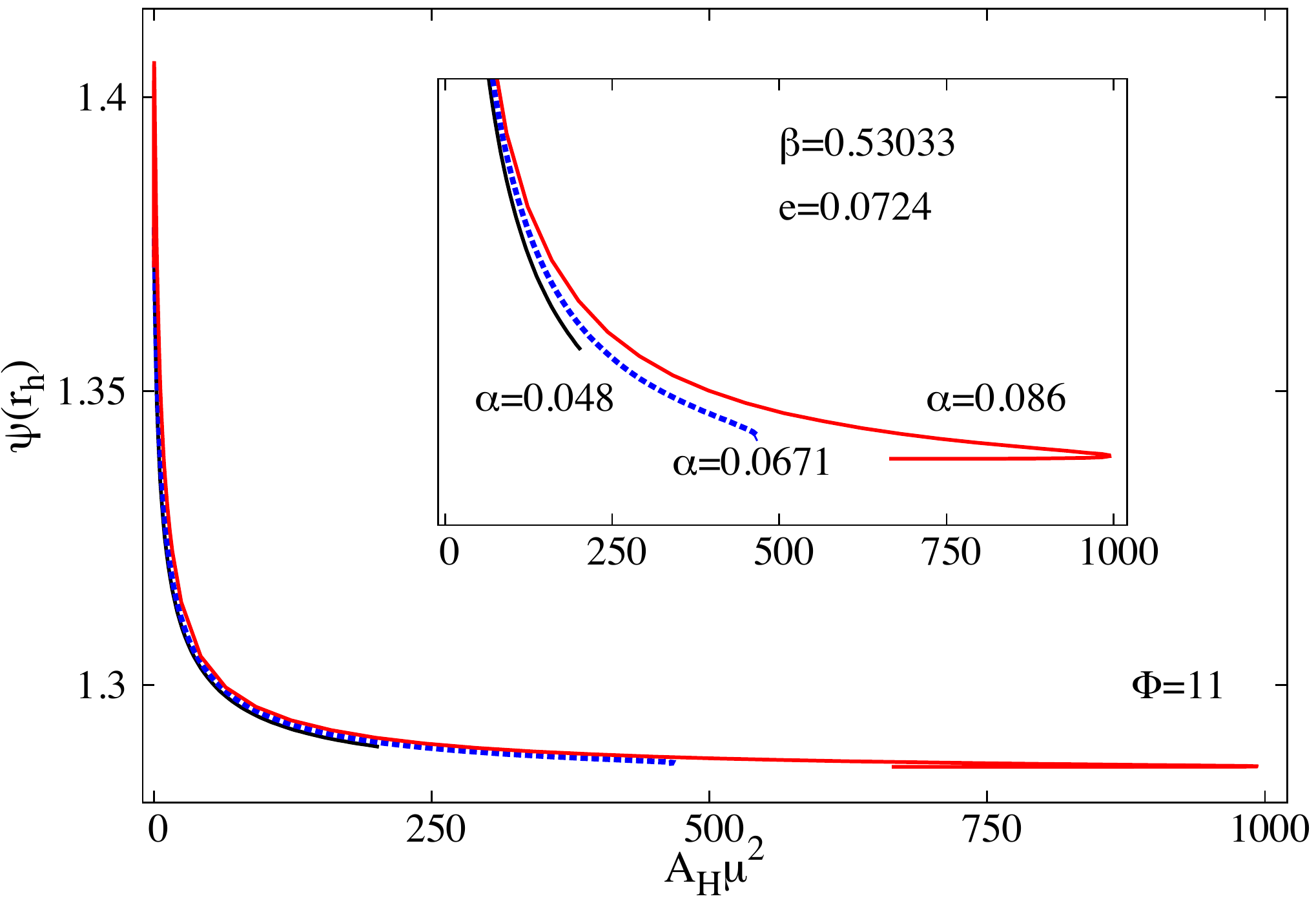}} 
	\resizebox{9cm}{7cm}{\includegraphics{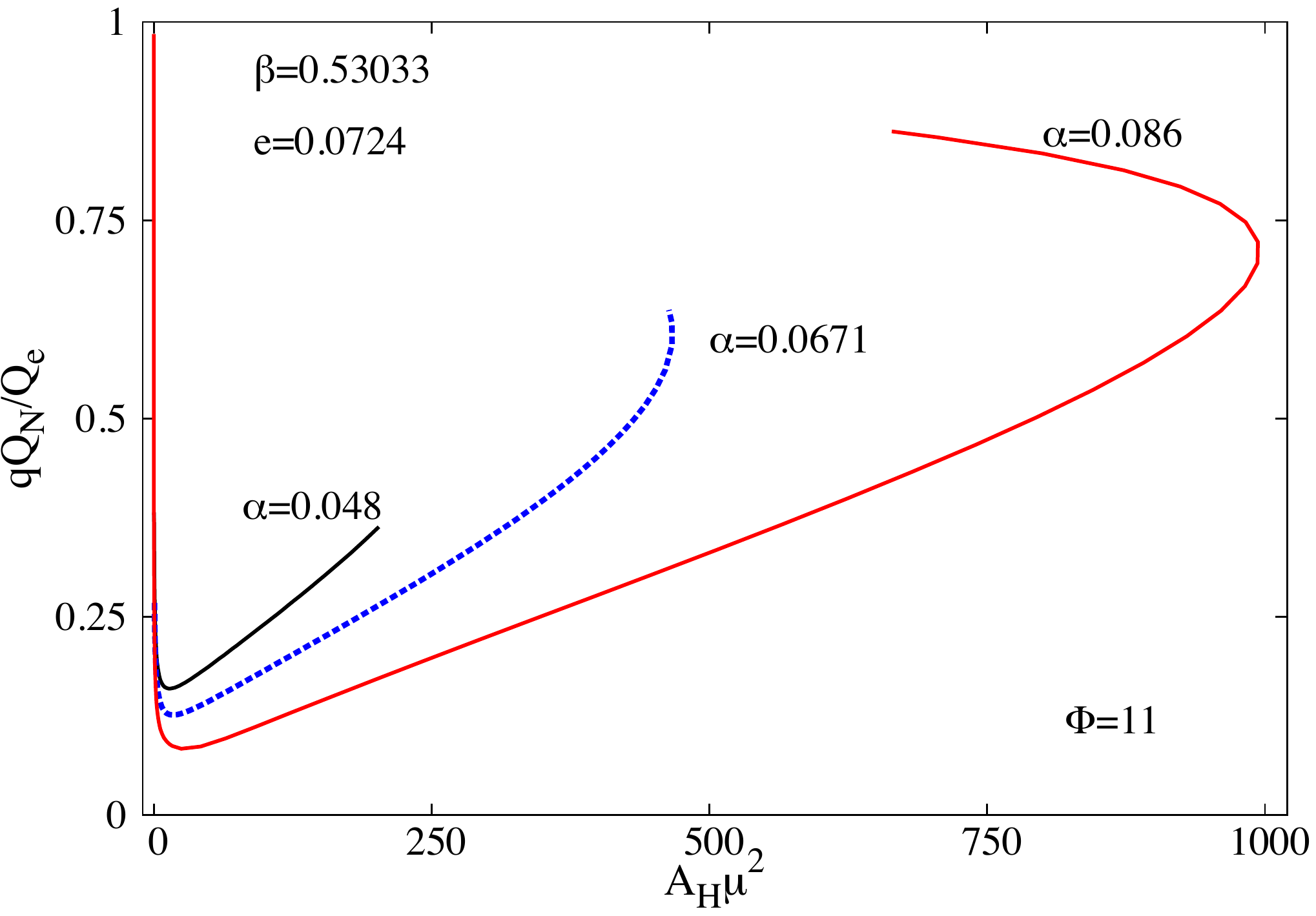}}  
\hss}
\caption{\small 
Value of the scalar field at the horizon $\psi_h$ (left panel) 
and the hairiness parameter
$q Q_N/Q_e$ (right panel) $vs.$ the horizon area, for the same solutions as in Fig.~\ref{gravity1}.
}
\label{gravity2}
\end{figure}

Fig.~\ref{gravity1} exhibits the behaviour of the mass and electric charge of the hairy BHs in terms of their horizon area, for illustrative values of $\beta,e$ and a sample of values of $\alpha$. Fixing these parameters, the solutions with a fixed chemical potential $\Phi$ 
($i.e.$ in a grand canonical ensemble)
exist from an arbitrary small size up to a maximal BH size, as specified by the event horizon area $A_H$. This defines the \textit{fundamental branch}. Along this fundamental branch, both the mass and the electric charge increase with $A_H$.
At the same time, the value of the scalar field at the horizon decreases - Fig.~\ref{gravity2} (left panel). 
As $A_H\to A_H^{(\rm max)}$, a secondary branch emerges, with a backbending in $A_H$. 
Thereafter, the numerics becomes increasingly challenging,
the scalar field
being confined in a region close to the horizon.
We suspect there may exist a spiraling structure, with  
extra branches and a critical central configuration.

 {\small \hspace*{3.cm}{\it  } }

In Fig.~\ref{gravity2} we can appreciate that neither the scalar field ever trivializes, nor the hairiness parameter $h=qQ_N/Q_e$, defined in (\ref{ph}), ever vanishes. This corroborates that these solutions have a mass gap with respect to the bald RN BHs and do not bifurcate from them.

 {\small \hspace*{3.cm}{\it  } }

One can take different perspectives on these hairy BH solutions. A first perspective is that any gauged boson star solution with the $Q$-ball potential appears to possess a BH generalization. That is, one can place a BH horizon within this gravitating soliton, under the resonance condition. 
Given the parameter 
$(\alpha,\beta, e;\Phi)$,
the BHs are found
 by  slowly increasing from zero the value of  
$r_h$ in the metric ansatz (\ref{metric}).
It is worth pointing out that the solitonic limit is smooth for the functions
$\delta(r)$, $\psi(r)$ only, 
while,  as $r_h\to 0$, the inner boundary behavior of 
$N(r)$  
           and
$V(r)$ 
jump from $0$ to $1$
   and  
 $0$ to $V_0\neq 0$, respectively.
Nonetheless, various global quantities like $M$, $Q_e$ and $\Phi$
are continuous as the soliton limit is approached.

A second perspective is that any $Q$-cloud on Schwarzschild/RN can be made backreact. These test field configurations arise for $\alpha=0$ and a given $r_h>0$.
The self-gravitating generalizations are found by slowly increasing the parameter $\alpha$.
We have found that
given $(\beta, e;r_h,\Phi)$,
the solutions exist up to maximally value of $\alpha$.

\begin{figure}[h!]
\hbox to\linewidth{\hss%
\resizebox{9cm}{7cm}{\includegraphics{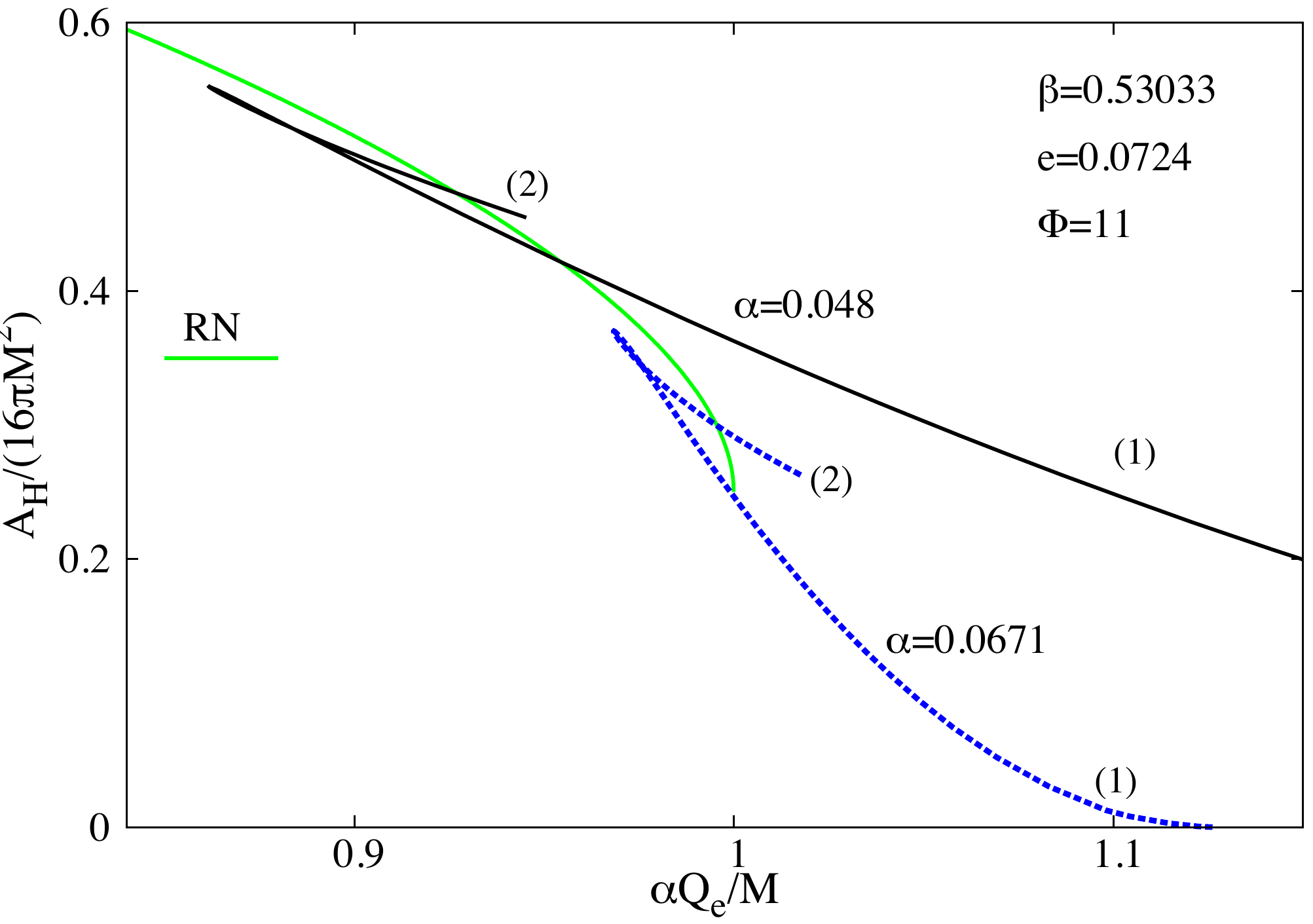}}
	\resizebox{9cm}{7cm}{\includegraphics{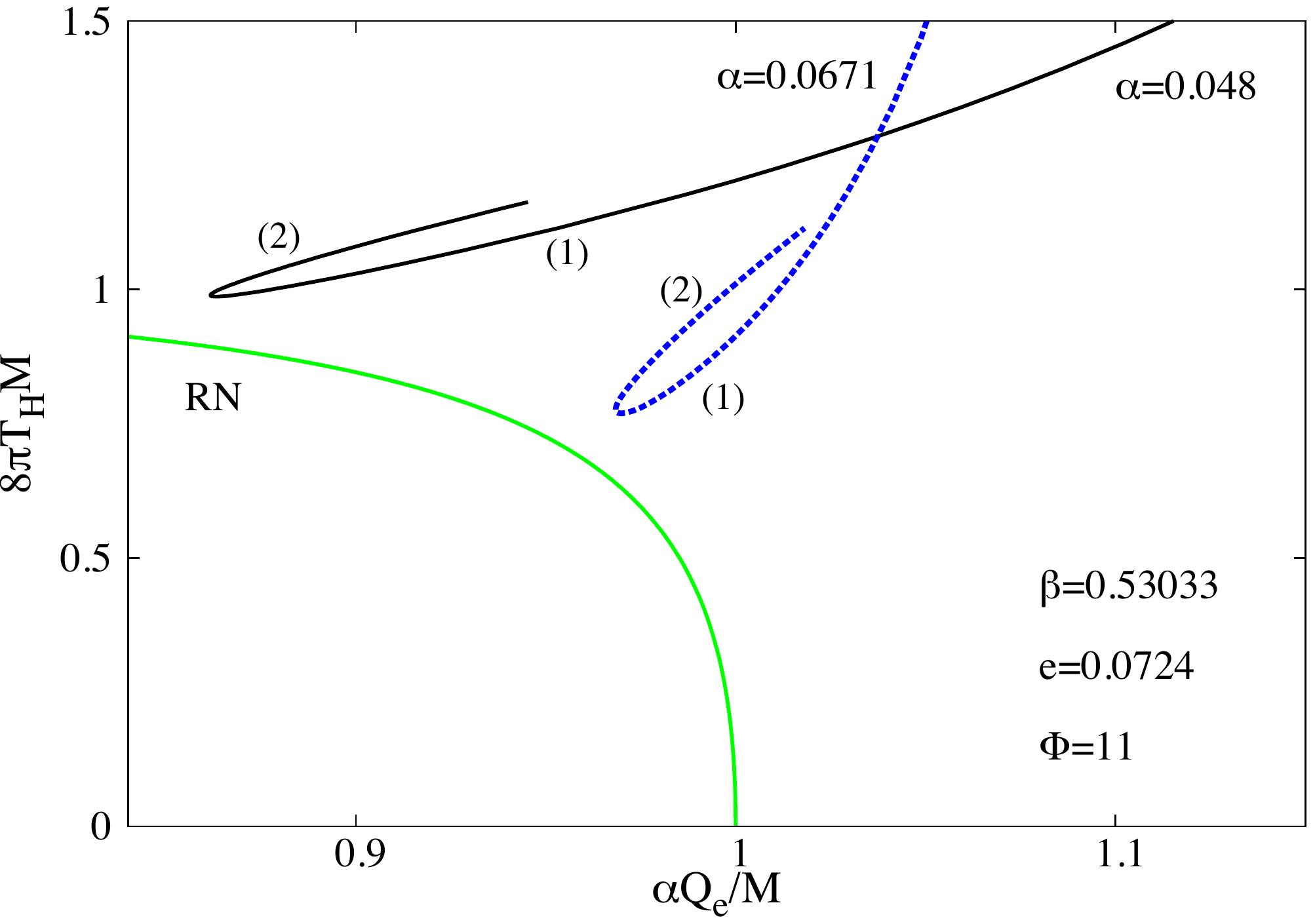}}  
\hss}
\caption{\small 
Reduced area (left panel) and temperature (right panel) of both $Q$-hairy BHs and the RN BH in terms of their charge to mass ratio for fixed $(\beta,e,\Phi)$ and three sample values of $\alpha$. The $Q$-hairy BHs can become entropically favoured sufficiently close to extremality and there is always a gap with the RN BH.
}
\label{gravity3}
\end{figure}

Finally, perhaps the most surprising perspective is that in the Einstein-Maxwell-scalar model (\ref{action}) the RN BH is not the unique spherically symmetric charged BH. Indeed, for the same charge to mass ratio there may exist also $Q$-hairy BH solutions, depending on the value of the remaining parameters. Moreover the hairy BHs can have larger area and hence entropically favoured. This is illustrated in Fig.~\ref{gravity3}. Fixing $\beta,e$ and the chemical potential $\Phi$, the hairy BHs for fixed $\alpha$ have a two branch structure. The fundamental branch connects to the solitonic limit $A_H\rightarrow 0$; after a backbending there is a second branch of larger area. One can see that, for the two values of $\alpha$, sufficiently close to extremality the hairy BHs have a larger area than the corresponding RN BH with the same charge to mass ratio. Moreover, there is a region where the non-uniqueness is three-fold: there are two hairy BHs and a RN BH with the same charge to mass ratio. In this sense, the situation resembles that recently reported in~\cite{Blazquez-Salcedo:2020nhs}. The right panel of Fig.~\ref{gravity3}, moreover, shows both that the temperature of the $Q$-hairy BHs never vanishes and also the gap between RN and the hairy solutions.

 {\small \hspace*{3.cm}{\it  } }

\section{Further remarks}
\label{sec7}

In this work we have shown that, 
contrary to common belief based in particular on the no-hair theorem by Mayo and Bekenstein~\cite{Mayo:1996mv}, a gauged scalar field minimally coupled to electro-vacuum can give rise to BH hair, as long as sufficient self-interactions 
are allowed.\footnote{Boson shells harbouring BHs 
with charged scalar hair
were considered in 
Ref. \cite{Kleihaus:2009kr}.
 However, those solutions require a $V -$shaped scalar potential which
is not of the form (\ref{potential}), and they possess rather different properties
as compared to the case in this work.
}
Observe, however, that the type of self-interactions necessary need not violate the weak or the dominant energy conditions; in fact, the examples herein use the common and physical $Q$-ball type potential. The obtained BHs can be interpreted as extensions of Schwarzschild or RN BHs with gauged, resonant, scalar hair. Their existence was anticipated by the results in~\cite{Hong:2019mcj}.
 Together with the results for synchronised scalar hair around Kerr~\cite{Herdeiro:2014goa} and Kerr-Newman BHs~\cite{Delgado:2016jxq}, the result herein establishes that all basic electro-vacuum solutions of GR allow some sort of extensions with BH hair, \textit{still within GR, with a minimally coupled scalar field}, albeit with different idiosyncrasies in each case. In this sense, BHs allow hair as a rule, rather than as an exception.

Perhaps the most intriguing apect of these new hairy BH solutions is why the solutions seem to require self-interactions of the scalar field, unlike the case of Myers-Perry BHs, described in the introduction. Further insight into this need would be desirable. Another issue concerns dynamical/stability properties of these solutions, which is an interesting direction of further research.

Also, let us mention that we have not fully explored the domain of existence of these $Q$-hairy BHs,
although a systematic study of the solutions seems possible.
In this direction, we have confirmed that besides the simple zero node (for the scalar field) solutions herein, there are also  solutions with nodes, corresponding to excited states. Furthermore, we predict the existence of similar hairy BHs for a gauged Proca field. Solitonic solutions with a gauged Proca field  have been discussed in $e.g.$~\cite{Garcia:2016ldc}.

Finally, we remark that BH solutions of
the Einstein-Maxwell-gauged scalar field model exist for $AdS$ asymptotics, providing the gravity duals of $s-wave$
superconductors \cite{Gubser:2008px}.
The main difference $w.r.t.$ the asymptotically flat case is that
those solutions emerge as perturbations around a RN-$AdS$ background. Thus, the nonlinearities of the scalar field potential
play no key role in this context.


\section*{Acknowlegements}

This  work  is  supported  by  the Center  for  Research  and  Development  in  Mathematics  and  Applications  (CIDMA)  through  the Portuguese Foundation for Science and Technology (FCT - Fundacao para a Ci\^encia e a Tecnologia), references UIDB/04106/2020 and UIDP/04106/2020 and by national funds (OE), through FCT, I.P., in the scope of the framework contract foreseen in the numbers 4, 5 and 6 of the article 23, of the Decree-Law 57/2016, of August 29, changed by Law 57/2017, of July 19.  We acknowledge support  from  the  projects  PTDC/FIS-OUT/28407/2017  and  CERN/FIS-PAR/0027/2019.   This work has further been supported by the European Union’s Horizon 2020 research and innovation (RISE) programme H2020-MSCA-RISE-2017 Grant No. FunFiCO-777740.  The authors would like to acknowledge networking support by the COST Action CA16104.

	


\end{document}